\documentclass[twocolumn,showpacs,aps,prl]{revtex4}
\usepackage{amssymb}
\usepackage{graphicx}
\usepackage{epstopdf}
\usepackage{dcolumn}
\usepackage{bm}
\usepackage{color}

\begin{document}

\title{Quantized magneto-thermoelectric transport in low-dimensional junctions}
\author{S.~E.~Shafranjuk}
\affiliation{Department of Physics and Astronomy, Northwestern University, Evanston, IL 60208, USA}

\date{\today }


\begin{abstract}
Quantization of the  magneto-thermoelectric transport is studied when an external d.c. magnetic field is applied to the C/N-knot formed as crossing between a narrow stripe of conducting atomic monolayer C on the one hand and metal stripe N on the other hand. The temperature gradient in C is created by injecting the non-equilibrium electrons, holes and phonons from the heater H thereby directing them toward the C/N-knot. A non-linear coupling between electron states of the C/N-knot counter electrodes causes splitting of the heat flow into several fractions owing to the Lorentz force acting in the C/N-knot vicinity, thereby inducing the magneto-thermoelectric current in N whereas the phonons pass and propagate along C further ahead. The heat flow  along C generates a transversal electric current in N showing a series of maximums when dimensions of the Landau orbits and the C/N-knot match each other. It allows observing the interplay between the quantum Hall effect and the spatial quantization.
\end{abstract}

\maketitle

Study of the magneto-thermoelectric transport in the low-dimensional conductors improves our knowledge about their nature and motivates the further work~\cite{Kane-Fisher,Shafr-TEG0}. Typical examples are the atomic monolayers (AM) such as graphene and transition metal dichalcogenides (TMDC) \cite{Supplement}. 
The electron transport in AMs is governed, e.g., by applying the gate voltages. It allows changing the charge carrier concentration and the magnitude of electrical conductance in wide limits~\cite{Wang}. Alternatively, electric  transport is controlled with the d.c. magnetic field.  At the low temperatures, it causes the quantum Hall effect which is manifested as quantization of the electrical conductance of the narrow stripe giving $\sigma =I_{ch}/I_{Hall}=\nu e^{2}/h$ where $\nu $ is the integer number. Furthermore, quantization of transverse electron motion occurs owing to spatial confinement of the electron states inside the narrow stripes. The above phenomena are utilized in the low-dimensional elements of electronic circuits, and also during the thermoelectric cooling and energy generation~\cite{TEG-book,Shafr-TEG}.

\begin{figure}
\includegraphics[width = 3.5 in]{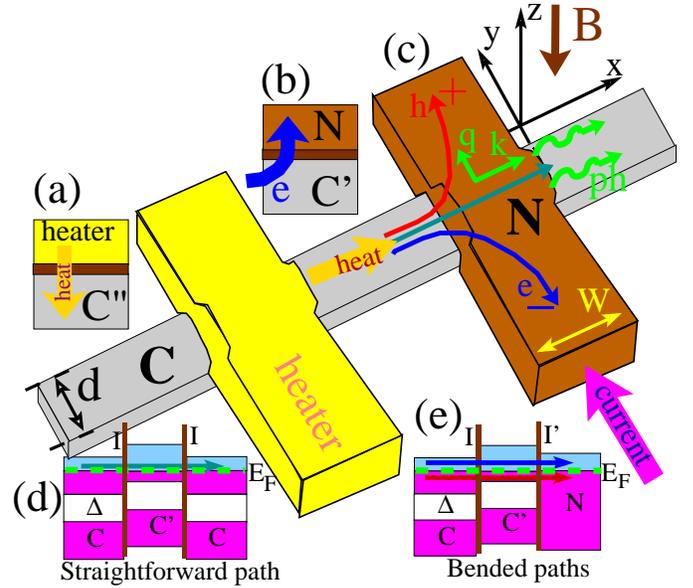}
\caption{{Color online. (a)~The injection of heat into the C$^{\prime \prime}$-section under the heater. (b)~Electron tunneling from the C$^{\prime}$-section into the stripe of metal  N.  (c)~Main figure: Splitting the flow of heat in the C/N-knot when a finite d.c. magnetic field $\mathbf{B}$ is applied perpendicular the C/N-knot plane. (d) and (e) represent the two different energy diagrams of the C/N-knot, which correspond to distinct trajectories of the electrons C/C$^{\prime}$/C (green arrow) and C/C$^{\prime}$/N (blue and red arrows).}} 
\label{fig1} 
\end{figure}
\begin{figure}
\includegraphics[width = 3.5 in]{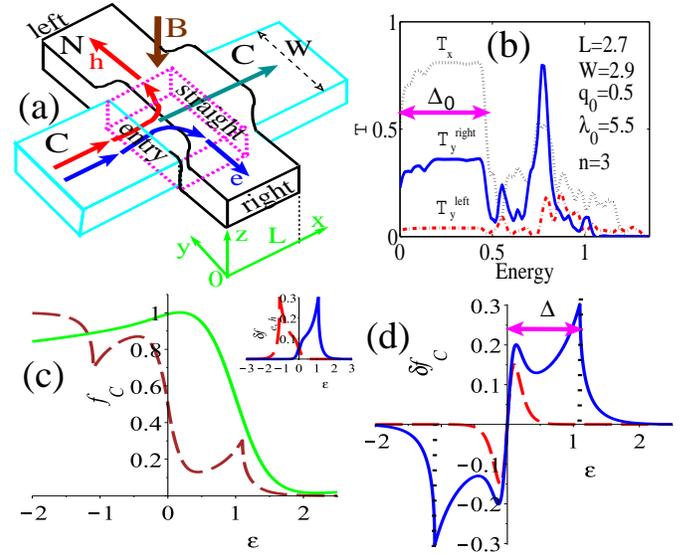}
\caption{Color online. (a)~The C/N-knot model for computing of the transmission probabilities ${{T_{y}}}$. (b)~The transmission probabilities for an electron emerging from C into the C$^{\prime }$-section along the {\it \^{x}}-direction, $T_{x}$ (doted black curve), turning to the right, ${T_{y}^{\rm right}}$ (solid blue curve), and turning to the left, ${T_{y}^{\rm left}}$ (dashed red curve) computed for geometry (a) by solving the non-linear boundary conditions.  (c)~The non-equilibrium electron distribution function $f_C (\varepsilon )$ in the C section. The brown curve is  $f_C (\varepsilon )$ which is altered only due to the phonon drag whereas the green curve is $f_C (\varepsilon )$  which is altered only due to the thermal injection of electron and holes (see parameters in text). Inset shows the phonon drag - induced non-equilibrium deviation of the distribution function $\delta f_{\rm e,h}$ of electrons (blue curve) and holes (red curve) in the C section for the same parameters. (d)~Driving factor of the transversal magnetoelectric current computed assuming that the quantum dot bottom $\Delta_0$ is set at $\Delta_0 = 0$ whereas the minigap is $\Delta = 1$ (the units are explained in text). The red dashed curve shows the influence of only the non-equilibrium thermal injection, whereas the solid blue curve shows the combined influence of the non-equilibrium thermal injection and phonon drag.} 
\label{fig}
\end{figure}

In this Letter we focus on the other aspects of the problem concerning of filtering,  spatial separation, redirecting and conversion the different components of the thermal and electric transport. Such the phenomena take place when the d.c. magnetic field  $\mathbf{B}=\{0,0,B_{z}\}$ acts on the crossing of a narrow AM stripe C with the metal stripe N. We examine the microscopic mechanism of controllable redirection and convertion one type of transport to another taking place in such the crossing (C/N-knot), where the electric and heat currents are separated from each other as shown in Fig.~\ref{fig1}. Thus, the C/N-knot represents a crossing of the narrow atomic monolayer material (AM) stripe C with the metal stripe N where the thermal flow components, i.e., electrons, holes, and phonons are rerouted in different directions as shown in Fig.~\ref{fig}a. The quantum dot is created on the C/N-knot as follows. Owing to the difference between work functions of the two materials at the C/N-interfaces, the Schottky barriers are formed. Furthermore, there are also barriers (marked in Figs.~1a,b,d,e as brown) created during the fabrication process due to presence of the atomic impurities localized at the interface, separating the open C-sections on the one hand and the C$^{\prime }$-section located underneath the metal stripe N on the other hand. Coupling of electron states of the C and N electrodes leads to several consequences: ({\it i}) modifies the Landau states, ({\it ii}) induces transmission of charge carriers between C and N and also ({\it iii}) shifts the energy profile by a finite value $\Delta_0$ serving as the bottom of the quantum dot. The heater H (shown in Fig.~\ref{fig1}a as yellow) injects the non-equilibrium phonons (ph), electrons (e), and holes (h) into the C-stripe, leading to the temperature difference $ \delta T^{\ast } = T^{\ast } - T_{C^{\prime }}$ arising between the section C$^{\prime \prime}$ beneath the heater on the left and C$^{\prime}$ beneath the metal N on the right. The effective local temperature $T^{\ast }$ in C$^{\prime \prime }$ much exceeds the steady state temperature $T_{\mathrm{C}}$ in the C-ends. According the Fourier law, the finite value of $ \delta T^{\ast } \neq 0$ initializes the flow of heat $Q=\Lambda \delta T^{\ast } $ along C between the sections C$^{\prime \prime}$ and  C$^{\prime}$ as shown in Fig.~\ref{fig1}. Here we introduced the thermal conductance $\Lambda =\Lambda _{\mathrm{ph}}+\Lambda _{\mathrm{e}}+\Lambda _{\mathrm{h}}$ consisting of the phonon ($\Lambda _{\mathrm{ph}}$), electron ($\Lambda _{\mathrm{e}}$), and hole ($\Lambda _{\mathrm{h}}$) components~\cite{Cahill}. The most interesting case is when numbers of the excited electrons and holes are equal to each other and they, jointly with phonons, carry only the heat along C but not the electric current, i.e.,  $I_{\parallel }\equiv 0$. It happens for the case of electron-hole symmetry of the excitation spectrum. A more detailed review of the corresponding C-materials is given in Ref.~\cite{Supplement}. As the counter-electrode N, one can use a normal metal stripe, e.g., Pd, Ni or Nb. In the AMs, most of the heat is transferred at expense of the non-equilibrium phonons. Furthermore, the propagating phonons drag additional electrons and holes along C~\cite{Bailyn}. Directing the d.c. magnetic field   $\mathbf{B}$ perpendicularly the C/N-knot, one spatially separates and  filters the ph-, e-, and h-components of the heat flow. In the C/N-knot, the Lorentz force deflects the electrons and holes (see respectively the blue and red arrows in Fig.~\ref{fig1}c) from C to N  in opposite directions, because their charges $\pm e$ are opposite, whereas the neutral phonons propagate further ahead (green arrow in Fig.~\ref{fig1}c).

We describe the {\it non-linear coupling} between electron states of the C- and N-stripes by introducing the partial transmision probabilities $T_{x,y}$. Let us derive the boundary conditions (BC) for the electron wavefunctions of the C/N-knot influenced by the external d.c. magnetic field as shown on Fig.~\ref{fig}a. First, we assume that after emerging from the open left C section, an electron then enters the C/N-knot (see Fig.~\ref{fig}a). For an electron passing through the C/N-knot, there are three possible further trajectories: ({\it a}) an electron is deflected from the straight trajectory toward the right wing of N (blue arrow in Fig.~\ref{fig}a); ({\it b}) it traverses the C/N-knot toward the open C-section on the opposite side further ahead (green arrow in Fig.~\ref{fig}a); ({\it c}) an electron, instead to be deflected toward the right wing of N, `wrongly' deviates from C toward the left wing of N. In the same figure, the red arrow depicts the hole trajectory. We emphasize that only the \textit{e}- and \textit{h}- deviated trajectories (see the blue and red lines in Fig.~\ref{fig1}c) are contributing to transverse electric current generated in N. Aforementioned processes effectively are represented as tunneling through the asymmetric C/C'/N-junction the energy diagram of whose is shown in Fig.~\ref{fig1}e.  The green arrow directed ahead as shown in Fig.~\ref{fig1}c depicts another type of trajectory. The latter process is represented as a tunneling via the symmetric C/C'/C junction (see the energy diagram in Fig.~\ref{fig1}d). Such the process gives no contribution into the electric current in N. 

The transverse magneto-thermoelectric current  $I_{\perp }$ is computed assuming that the non-equilibrium electrons and holes in C are excited by two mechanisms: ({\it i})~due to the phonon drag of the electrons and holes along C, and ({\it ii})~via the thermal injection of the non-equilibrum quasiparticles from the heat source H into C (see Fig.~\ref{fig}a). More detailed description of the mechanisms ({\it i}) and ({\it ii}) is given in Ref.~\cite{Supplement}. Furthermore, we suppose that the C/N-interface transparency is high, thereby ensuring that the transmission probability  $\overrightarrow{\overleftarrow{T_{y}}}$ through the C/C$^{\prime }$/N knot is considerable for electrons (and holes) emerging from section C, then traversing the section C$^{\prime }$, and being transmitted, e.g., in the left (or right) wing of N, provided $\overrightarrow{\overleftarrow{T_{y}}} \lesssim 1$.  The transmission probabilities $\overrightarrow{\overleftarrow{T}}_{y}$ are determined by the work functions of C and N. Simultaneously, the transparency $T_{ph}^{\mathrm{CN}}$ for phonons of the same C$^{\prime }$/N-interface is very low, i.e., $T_{ph}^{\mathrm{CN}}<<1$, though propagation of phonons inside C is ballistic. That happens because redirecting of the phonons from C to N is accompanied by a significant change of the phonon momentum $\Delta \mathbf{q}= \hat{\mathbf{y}}q_{\mathrm{N}} - \hat{\mathbf{x}}q_{\mathrm{C}}$ where $q_{\mathrm{N}}$ and $q_{\mathrm{C}}$ are the phonon momentum components along N and C respectively~\cite{Cahill}. Another reason why $T_{ph}^{\mathrm{CN}}<<1$, is that the phonon spectrums of C and N are very distinct. Therefore, for the sake of simplicity, we disregard the phonon transmission from C to N by setting $T_{ph}^{CN}\simeq 0$. On the one hand, we assume that phonons pass the knot and propagate along C further ahead. On the other hand, we will see that the Lorentz force reroutes electrons and holes propagating between C and the perpendicular stripe of the metal N. As it follows from the further calculations, the redirection happens because the Lorentz force pulls electrons to the right (blue arrow in Fig.~\ref{fig1}c) while simultaneously pushing the holes toward the left (red arrow in Fig.~\ref{fig}c). In this way, the heat flow (yellow arrow in Figs.~\ref{fig1}a,c) along C induces the electric current $I_{\perp } $ (magenta arrow) in N (brown).

Here we consider most spectacular case of the electron-hole symmetry in C assuming that the  energy gap $\Delta$ is relatively small.  The magnitude of the transversal electric current generated in N is found according the Landauer-B\"{u}ttiker formula%
\begin{eqnarray}
I_{\perp } &=&\frac{4e}{h}%
\int_{0}^{\infty }d\varepsilon M_{\min }\left( \varepsilon \right)  \left( \overleftarrow{T}_{y}\left( \varepsilon \right) -%
\overrightarrow{T}_{y}\left( \varepsilon \right) \right)  \nonumber
\\
&&\times \left(f_{C}\left( \varepsilon -V_{\rm H} \right) -f_{N}\left( \varepsilon \right) \right)
\label{current}
\end{eqnarray}%
where $M_{\min }\left( \varepsilon \right) $ is the minimum number of quantum channels per energy interval, $\overleftarrow{\overrightarrow{T}}_{y}\left( \varepsilon \right) $ is the energy-dependent transmission probability through the C/C$^{\prime }$/N knot for an electron emerging from section C, traversing the section C$^{\prime }$, and getting into the left (or right) wing of N, $V_{\rm H}$ is the thermal Hall voltage induced by the d.c. field across C as explained in Ref.~\cite{Supplement}. When the excitation spectrum in C is non-symmetric, and one type of charge carriers (either electrons or holes) prevails, in addition to the heat flow, there is also a finite thermoelectric current, also flowing along C. 
\begin{figure}
\includegraphics[width = 3.5 in]{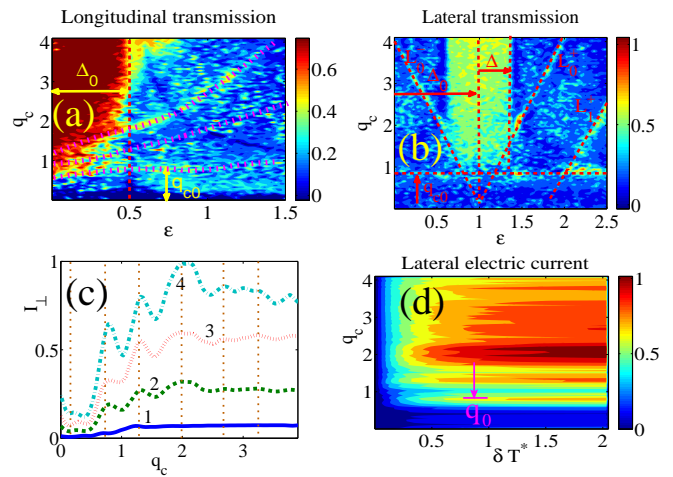}
\caption{Color online. (a)~Longitudinal transmission probability through the C/N-knot versus the electron energy $\varepsilon $ (in units $2\Delta_0$) and the cyclotron wave number $q_c = \sqrt{eB_z/\hbar }$ (in units $2\Delta_0 /v$) computed assuming $\Delta = 0$ and $2\Delta_0 =1$ (see units and other parameters in text). The dotted lines indicate geometrical resonances originating from matching between the Landau orbits and the C/N-knot dimensions. Notice the features related to $\Delta_0 $ and $q_{c0} = \sqrt{q_c^2+q_0^2}$ where $q_0 = \sqrt{2 m  \omega_0/\hbar}$. The parameters characterize the bottom energy bands  of C$^{\prime }$. (b)~The transversal transmission probability for $2\Delta_0 = 1$ and $\Delta = 0.4$. The dashed lines indicate resonances $L_{0,1}^{\pm}$ related to the Landau and spatial quantization. Notice also features related to the bottom  energy $\Delta_0$, the energy gap $\Delta$, and the cyclotron wave number $q_c$. (c)~The transversal magneto-thermoelectric current $I_{\perp}$ (in units $2\Delta_0 (e/h)$) generated in N by the heat flow along C as a function of the cyclotron wave number $q_c$ for different temperature deviations $\delta T^*$ in C (see text).  (d)~Corresponding contour plot of the steady state $I_{\perp}$ for the same parameters of the C/N-knot as before.} 
\label{fig3} 
\end{figure}

The \textit{transmission probability} $\overleftarrow{\overrightarrow{T}}_{y}\left( \varepsilon \right) $ through the C/N-knot is computed using the model depicted in Fig.~1c and Fig.~\ref{fig}a. As the trial electron wavefunction $\Psi_{\mathrm{C^\prime }} \left( x,y\right)$, we use a combination of two-dimensional waves formed inside the rectangular box situated on the C$^\prime $-section underneath the metal N as shown in Fig.~\ref{fig}a. The box represents the rectangular 2D quantum dot whose bottom is at the energy value $\varepsilon = \Delta_0$. For the sake of simplicity, we disregard the particle's chirality~\cite{Geim-Graph}.  The value of $\Delta_0$ is set owing to coupling between the atomic orbitals of C and N. We assume that an electron with the momentum ${\bf p}=\{k,q\}$ enters the C/N-knot on the left side from C to C$^\prime $ at $x=0$ and $0<y<W$. The electron leaves the C/N-knot toward N either on the right ($y=0$, $0<x<L$), on the left ($y = W$, $0<x<L$), or propagates toward C further ahead ($x=L$, $0<y<W$) where $L$ is the metal stripe width. While traversing the C/N-knot, the electrons (holes) acquire a transversal momentum component $q = \pm \left( \left( e/c\right) B_z y + t^{2}g/v\right) $ where the former term is due to magnetic field whereas the last term is caused by the C/N-coupling whose energy is $t^{2}g$. Here $t$ is the matrix element of tunneling between C and N, $g$ is the electron density of states in the metal stripe N, $v$ is the Fermi velocity in C. The confinement effect inside the C-stripe is modelled with introducing the spreading of parabolic potential in the $\hat y$-direction $U\left( y\right) = m \omega _{0}^{2}y^{2}/2$ where $m$ is the electron mass and $\omega _{0}$ characterizes the steepness of the confinement potential. Then the dispersion law of an electron inside C is 
\begin{equation}
E_{n,k} = E_{s}+\hbar \omega _{c0}\left( n+\frac{1}{2}\right) +\frac{\hbar ^{2}k^{2}%
}{2m}\frac{\omega _{0}^{2}}{\omega _{c0}^{2}}.
\label{electr-energy}
\end{equation}
where $n$ is a non-negative integer, $\omega_c = eB/m_e$ is the cyclotron frequency, and $\omega _{c0} = \sqrt{\omega _{c}^{2}+\omega _{0}^{2}}$. From Eq.~(\ref{electr-energy}), it follows that both, the confinement of electron states and the d.c. magnetic field, cause minibands in the electron excitation spectrum of the C-stripe. For the rectangular geometry of the C/N-knot  shown in Fig.~\ref{fig}a, the \textit{trial electron wavefunction} takes the form
\begin{equation}
\Psi_{\mathrm{C^\prime }} \left( x,y\right) =\left( \text{$\alpha $}%
_{x}e^{ikx}+\text{$\beta $}_{x}e^{-ikx}\right) \left( \text{$\alpha $}%
_{y}u_{n}\left( \zeta_{+}\right) +\text{$\beta $}_{y}u_{n}\left(
\zeta_{-}\right) \right)  \label{psi}
\end{equation}%
where we introduced auxiliary functions $u_{n}\left( \zeta \right) =\exp \{ -\zeta^{2}/2\} H_{n}\left( \zeta\right) $, $H_n(\zeta )$ is the Hermite polynomial, $\zeta_{\pm }=\zeta\left( y,\pm k\right)  = q_{0}y\mp \lambda _{c0}k $. Here $q_{0}=\sqrt{m\omega _{c0}/\hbar }$, $\lambda _{c0}=(\omega_{c}/\omega _{c0}) ^{2}\lambda ^{2}q_{0}$ and $\lambda = \sqrt{\hbar /eB}$ is the magnetic length. Furthermore, at $B=0$, $\pi/q_0 = W^*$ is the effective width of C-stripe.  For the magnetic flux density $B=16\rm {T}$ one obtains $\lambda =\sqrt{\hbar /eB}\simeq 6.4$~nm, $\hbar \omega _{c}=2$~meV, $\hbar \omega _{0}=5$~meV, $\hbar \omega _{c0}\simeq 5.3$~meV, and the effective  width of the stripe is $W^*=\pi/q_{0}\simeq 75$~nm, $U_{0}=\Delta _{0}\simeq 5.5$~meV where $\Delta_{0}=E_{s}+\hbar \omega _{c0}/2$ is the bottom energy.

Outside the C/N-knot region, the trial wavefunctions are taken in the form of plane waves. For an electron which is propagating along the C-stripe, at $x<0$ and $y=0$ we use
\begin{equation}
\Psi _{C}\left( x,y=0\right) =r_{x}e^{-ik_{1}x}+e^{ik_{1}x}
\label{wave-C}
\end{equation}
whereas for an electron inside the N-stripe $x=0$  we use
\begin{equation}
\Psi _{N}\left( x=0,y\right) =t_{y}^{R(L)}e^{iq_{3}(y\pm W/2)}
\label{wave-N}
\end{equation}%
where the upper (lower)  signs correspond to $y>W$ ($y<-W$). In the above formulas (\ref{wave-C}), (\ref{wave-N}) we disregarded dependence on $y$,  which enters into the coefficients $r_x$ and $t_y^{R(L)}$. The transmission probabilities are defined, e.g.,  as $\overrightarrow{\overleftarrow{T}}_{y}(\varepsilon )=\left\vert t_{y}^{R(L)}\right\vert ^{2}$. The energy dependence of $k$ is then obtained in the form $k (\varepsilon) = (\omega _{c0}/\omega_{0})(\sqrt{2m\Delta _{n}}/\hbar )\sqrt{(\varepsilon -\Delta _{n})/\Delta
_{n}}$ where $\Delta _{n}=E_{s}+\hbar \omega _{c0}\left( n+1/2\right) $. The boundary conditions (BC) yield the system of 8 non-linear transcendental equations for the 8 coefficients entering the trial wavefunction (\ref{psi}). Technically, non-linear coupling between the electron states in the C- and N-stripes  originates from the products like $\propto \alpha_x \alpha_y$, $\alpha_x \beta_y$, $\beta_x \alpha_y$,  which are present in the boundary conditions. The non-linear BC equations are solved numerically.  The obtained solutions allow to compute the transmission probabilities $\overleftarrow{\overrightarrow{T}}_{y}(\varepsilon )$ which enter in Eq.~(\ref{current}). Our simple model assumes that the propagation of particles within the C/N-knot is ballistic.  The model also can be extended to include the roughness of substrate and interfaces as described, e.g., in Ref.~\cite{Perebei}.

The electric current $I_{\perp }$ is induced in the N-stripe owing the thermal injection~\cite{Shafr-TEG0}  and the phonon drag~\cite{Bailyn}. Thereby, $I_{\perp }$ depends on the non-equilibrium distribution functions of the electrons $f_{\mathrm{C}}$ and phonons $N_{\mathrm{C}}$ inside the C/N-knot. The functions $f_{\rm C}$ and $N_{\mathrm{C}}$ are found as solutions of the kinetic equations%
\begin{eqnarray}
\dot{f}_{\mathrm{C}} = \mathcal{L}_{th} + \mathcal{L}_{ep}  \label{kin-el} \\
\dot{N}_{\mathrm{C}} = \mathcal{P}_{th} + \mathcal{P}_{pe}  \label{kin-eq}
\end{eqnarray}%
where $\mathcal{L}_{th}$ and $\mathcal{P}_{th} $ describe respectively the thermal injection of the non-equilibrium electrons, holes, and phonons from the heat source H into the C$^{\prime \prime}$ section as shown in Fig.~\ref{fig1}a, $\mathcal{L}_{ep}$ is the electron-phonon collision integral, whereas $\mathcal{P}_{pe}$ is the phonon-electron collision integral. In Eqs.~(\ref{kin-el}), (\ref{kin-eq}) we have disregarded other processes having lesser importance. 

The {\it non-linear coupling} between the electron states of counter electrodes in the C/N-knot causes the leakage of the electrons and holes from the C-stripe to N as illustrated in Fig.~\ref{fig}. The corresponding numeric results are presented in Figs.~2b-d. In Fig.~\ref{fig}b we plot the transmission probabilities for an electron incoming from C into the C$^{\prime }$-section along the \^{x}-direction, $T_{x}$ (dotted black curve), turning to the right, $\overrightarrow{T}_{y}$ (solid blue curve), and turning to the left, $\overleftarrow{T}_{y}$ (dashed red curve) obtained as solutions of the non-linear boundary conditions. The probabilities of electron transmission $\overrightarrow{\overleftarrow{T}}_{y}$ and $T_{x}$ contain series of peaks related to the geometrical resonances arising when the dimensions of Landau orbit with index $n=3$ and the C/N-knot dimensions coincide with each other. Notice the buildup of the transmission probabilities at energies $0 < \varepsilon < \Delta_0$ where $\Delta_0 $ is the bottom energy of the C/N-knot. The buildup arises when the effective width $W^* = \pi/ q_0$ of the C-stripe is finite. By other words, the maximums occur when position of the Landau level $n=3$ matches with the localized energy levels originating from the spatial quantization. Thus, stronger transversal deviation of electrons and holes traversing through the C/N-knot occurs at certain values of the electron energy $\varepsilon $ and magnetic field $B_{z}$. To illustrate the difference between the two mechanisms of creating the non-equilibrium excitations (i.e., the phonon drag and thermal injection) in C, we set  either $\delta f^{\mathrm{pd}}\equiv 0$ while holding $\delta f^{\mathrm{th}}\neq 0$ or, otherwise, $\delta f^{\mathrm{th}}\equiv 0$ while $\delta f^{\mathrm{pd}}\neq 0$. The calculation parameters depend on properties of the C-  and  N-materials. Furthermore, certain parameters,  like the magnetic length $\lambda $ and the cyclotron frequency $\omega _{c} $ also depend on whether the  electron excitations are the chiral fermions or not. Therefore it is convenient to operate with material-independent dimensionless units. We use the following parameters $\tau _{\mathrm{ep}}=10^{-12}$~s, $\sigma _{0}=0.2$, $N_{0}=1$, $\Delta =1$, $T_{\mathrm{C}}^{\prime }=0.1$ is the temperature in C$^{\prime }$-section, $s=0.02$ is the sound velocity, $v=1$. The dimensionless length units correspond to $v/2\Delta_0 $ whereas the wavevectors are in the units of $2\Delta_0/v $ where both $\Delta_0$ and $v$ are material-dependent. For the sake of convenience, in Fig.~\ref{fig}b and below, the parameters with energy and temperature dimensions are expressed in units of the semiconducting gap $\Delta $ in C (we set $\hbar =1$). Parameters with the wave number dimensions (like $k$, $q_{0}$, $q_{c}$, etc.) are expressed in units $\Delta /v$ whereas the length parameters (like $L$, $W$, $\lambda $, $\lambda _{\mathrm{c0}}$, etc.) are expressed in the units of $v/\Delta $. Then, e.g., for the C-stripe width $W=75$~nm one gets $\Delta =2\hbar v/W=0.4$~eV, $\Delta /v=7.5\times 10^{8}$~m$^{-1}$ and $v/\Delta =1.3$~nm where we also use that $v=8.1\times 10^{5}$~m/s and $s=2.1 \times 10^{4}$~m/s. In Fig.~\ref{fig}c, the brown curve represents the distribution function of electrons in C, $f_C (\varepsilon )$, altered owing to only the phonon drag, whereas the green curve is $f_C (\varepsilon )$ altered only by thermal injection of the electron and holes (see parameters in text). Inset in Fig.~\ref{fig}c shows the phonon drag - induced non-equilibrium deviation of the distribution function $\delta f_{\mathrm{e,h}}$ of electrons (blue curve) and holes (red curve) in the C section for the same parameters as listed above. Driving factor $\delta f_{\mathrm{C}}=f_{\mathrm{C}}-f_{\mathrm{N}}=\delta f^{\mathrm{pd}}+\delta f^{\mathrm{th}}$ for the transversal magnetoelectric current $I_{\perp }$ at $U_{0}=0$ is plotted in Fig.~\ref{fig}d. Red dashed curve corresponds to influence of the non-equilibrium thermal injection only (for the moment we set $\delta f^{\mathrm{pd}}\equiv 0$) while the blue solid curve shows a combined influence of both, the non-equilibrium thermal injection and the phonon drag.

In Fig.~\ref{fig3} we show the obtained transmission probabilities $T_{x}\left( \varepsilon \right) $, $\overleftarrow{T_y}\left( \varepsilon \right) $ and the transversal electric current  $I_{\perp } $ versus magnetic field $B_{z}$ and temperature deviation $\delta T^{\ast }=T^{\ast } - T_{\mathrm{C}}$. The effective electron temperature $T^{\ast }$, as a characteristic of the non-equilibrium effect, is introduced as $T^{\ast } =  (\Omega_{\rm CN}/k_{\rm B})\bar{\varepsilon} _{el}$, where $\Omega_{\rm CN} = h W L$ is the C/N-knot volume, $h$ and $W$ are respectively the thickness and width of C, and $L$ is the width of N. The energy of non-equilibrium electrons per unit volume $\bar{\varepsilon} _{el}$ is given by Eq.~(12) of Ref.~\cite{Supplement}. Maximums of the electric current component $I^{(3)}_{\perp }$ in Fig.~\ref{fig3} arise from matching of the Landau orbit ($n=3$) with dimensions of the C/N-knot. In other words, they constitute an interplay between the Landau and localized levels arising in the C/N-knot. 
Figure~\ref{fig3} suggests that the basic C/N-knot parameters can be extracted straight from the magneto-thermoelectric characteristics. 
In Fig.~\ref{fig3}, we use $L = 0.6$,  $W = 0.67$, and $h=10^{-3}$, all expressed in the dimensionless units.  The bottom energy $\Delta_{0}$ of the C/N-knot and its confinement steepness wave number $q_{0}$ are respectively $\Delta_{0}=0.5$ and $q_{0}=0.5$. When the C-stripe is made of,  e.g.,  NbSe$_2$,  the dimensionless units listed for Fig.~\ref{fig3} correspond to the following parameters dependent on the physical properties of the material. The d.c. magnetic flux density (magnetic induction)  is $B=16$~T,  $\lambda =6.4$~nm, the cyclotron energy $\hbar \omega _{c}=1.85$~meV, the confinement steepness energy $\hbar \omega _{0}=5$~meV, $\hbar \omega _{c0}\simeq 5.3$~meV, the effective C-stripe width $W^* = \pi/q_{0} = 80$~nm, $v/2\Delta_0 = 130$~nm, $2\Delta_0/v = 7.6 \times 10^{6}$~m$^{-1}$, $\Delta _{0} = E_s+\hbar \omega_c/2 = 0.5$~meV.   Figure~\ref{fig3}a shows the longitudinal transmission probability $T_{x}\left( \varepsilon \right) $ through the C/N-knot versus the electron energy $\varepsilon $ and the cyclotron wave number $q_{c}=\sqrt{eB_{z}/\hbar }$ setting the energy gap $\Delta = 0$ and the bottom energy $2\Delta_0 = 1$.  In Fig.~\ref{fig3}b we present the transversal transmission probability $\overleftarrow{T_y} (\varepsilon )$ through the C/N-knot versus $\varepsilon $ (in units $2\Delta_0$) and $q_{c}$ in units $2\Delta_0/v$,  assuming that the energy gap is finite, $\Delta = 0.4$. In Fig.~\ref{fig3}c we plot the transversal magneto-thermoelectric current $I^{(3)}_{\perp }$ [in units $2\Delta_0 (e/h)$ by setting $\Delta = 0$ again] induced along N versus the cyclotron wave number $q_{c}$ (in units $2\Delta_0 /v$) for different effective temperature deviations $\delta T^{\ast }=0.08,0.8,1.4$, and 2 (in units $2\Delta_0/k_{\rm B} $) for curves 1-4 respectively. As an illustration, in Fig.~\ref{fig3}d, we also show the corresponding contour plot of $I^{(3)}_{\perp }$ for the same C/N-knot parameters as in the former Fig.~\ref{fig3}c. From Figs.~\ref{fig3}c,d one can see that $I^{(3)}_{\perp }$ is roughly proportional to the temperature deviation $\delta T^{\ast }$ in C. 

The obtained results suggest that the electrons (e) and holes (h),  having opposite electric charges and propagating in C under influence the Lorentz force (see Fig.~\ref{fig1}b) are diverted toward N in opposite directions. Thus, a finite field $\mathbf{B} \neq 0$ splits the heat flow into several components, thereby separating the electrons and holes from each other as shown in Fig.~\ref{fig1}a and causing a finite electric current $I_{\mathrm{e}} = I_{\mathrm{h}} = I\neq 0$ along N.
The magnitude of electric current induced in N is proportional to the heat flow in C and also depends on the d.c. magnetic field $\mathbf{B}$. The non-linear tunneling coupling between the electron states of the C- and N-stripes modifies the Landau states in the C/N-knot thereby introducing a difference between the magneto-thermoelectric transport in the C/N-knot and isolated C-stripe. 

Our results illustrate that the study of magneto-thermoelectric phenomena enable better understanding the microscopic transport mechanisms in the atomic monolayer materials and junctions, thereby opening many additional opportunities in a variety of the research and applications. By measuring resonances of $I_{\perp }$, one establishes key parameters of the AM junction and examines interplay between the Landau and spatial quantizations in conditions of the  thermal quantum Hall effect. It extends capabilities to determine the parameters of low-dimensional junctions on the nanoscale. In principle, the filtering of heat flow  in the C/N-knot subjected to the d.c. magnetic field can be used, e.g., to probe the electron-electron interaction and to potentially reveal other spectroscopic features like, e.g., the parameters of Luttinger liquid or exciton binding energy in narrow AM stripes. Furthermore, the phenomenon can be utilized for the thermoelectric cooling and generation of electricity~\cite{TEG-book,Shafr-TEG0}.

This work had been supported by the AFOSR grant FA9550-11-1-0311.

\section{Appendix}

\subsection{C-materials}
There are many monoatomic layer conductors regarded in this paper as C. One example is graphene \cite{Geim-Graph}, an extremely thin electric and thermal conductor \cite{Baland}, showing high carrier mobility \cite{Mayor}, and surprising properties of the molecular barrier \cite{Bunch,Nair}. Other 2D materials are TMDCs \cite{Wang}, transition metal oxides like titania- and perovskite-based oxides \cite{Osada,Ayari}, and graphene analogues such as boron nitride (BN) \cite{Dean,Pacile}. Atomic monolayers of TMDCs which display the wide range of interesting electronic, optical, mechanical, chemical and thermal properties, have a strong potential for a variety of applications. The knowledge obtained from graphene \cite{Wang} helps in the sample preparation, optical detection, transfer and manipulation of 2D materials, and in general understanding of 2D material physics.
Numeric study \cite{Kuc,Kobayashi,Li} suggests that the band structure of different TMDCs is similar in general features, which also is confirmed using a variety of spectroscopic tools \cite{Wang,Mak,Splendiani}. The MoX$_2$ and WX$_2$ compounds are semiconducting, whereas NbX$_2$ and TaX$_2$ are metallic \cite{Wang}, where X is a chalcogen (S, Se or Te). Direct excitonic transitions near the K-point remain relatively unchanged versus the layer number \cite{Wang,Splendiani}.

Currently, there is a strong interest to the monatomic materials with relatively {\it small energy gap} $\Delta_l \simeq 1-100$~meV. The interest is motivated by a variety of potential applications in the nanoelectronics, e.g., the THz sensors and spectral analyzers, lasers, high-speed logic, and superfast computers. Typically, the pristine monoatomic layer materials are either semiconducting (e.g., such TMDCs as MoX$_2$ and WX$_2$) or metallic (e.g., NbX$_2$ and TaX$_2$). The energy minigap $\Delta_l$ in the excitation spectrum of a monatomic metal is induced using one of the many approaches \cite{Lado}. For instance, a minigap can arise due to the interaction-driven electronic order \cite{Min,Araki,Semenoff} when a high magnetic field is applied; it also can result from the influence of substrates such as SiC \cite{Zhou}, \cite{Bostwick} or hexagonal boron nitride (BN), \cite{Giovannetti} although the lattice mismatch might cause rather complex effects \cite{Dean}.  Ref.~\cite{Park} describes the superlattice patterns on graphene created using the electron beam-induced deposition of adatoms \cite{Meyer}. The pattern periodicity is achieved as short as $\simeq 5$~nm. The triangular patterns with the lattice period $L_{\rm SL} \sim 10$~nm were formed for graphene on metal surfaces \cite{Marchini,Vazquez,Pan}. The semiconducting and metallic superlattices routinely are used for manipulating the electronic structure of materials. Superlattices are fabricated by placing the gate arrays on the graphene sheet. 

The scenario is similar for different monoatomic conductors. A simple model considers a "slow" periodic potential applied to the atomic monolayer thereby inducing {\it minibands in the electron excitation spectrum}. 
We assume that a periodic 1D potential $V(x)$, with period $L_{\rm SL}>>a$ ($a$ is the lattice constant) along the $x$-direction, is applied, e.g., to graphene. The dispersion law for an electron with momentum ${\bf p}=\{p_x,p_y\}$ is derived \cite{Park} in the form $E^l_{s}\left( \mathbf{p}\right) =\pm \hbar v\sqrt{p_{x}^{2}+\left\vert f_{l}\right\vert ^{2}p_{y}^{2}}+\pi \hbar v l /L_{\rm SL}$ where $v$ is the Fermi velocity, $f_{l}$ are the coefficients of the $V(x)$ expansion $\exp\{2i \int_0^x dx^{\prime } V(x^{\prime })/(\hbar v)\}=\sum_{l=-\infty }^{\infty }f_{l}\left[ V\right] e^{i \pi lx/L_{\rm SL}}$, $l$ is integer. Then one gets the minigaps $\Delta _{l}=\hbar v\left\vert f_{l}\right\vert p_{l,y}$ spaced with $\Delta E_{s}\left( \mathbf{p}\right) =\pi \hbar v /L_{\rm SL}$. The electron density of states in the C-stripe then contains series of Van Hove singularities associated with the minigaps $\Delta_n$. Such series of Van Hove spikes with $2\Delta_n \simeq 0.4-0.8$~meV was experimentally observed in graphene on the hexagonal boron nitride (hBN) substrate as reported in Ref.~\cite{Hunt}.  The equilibrium density of charge carriers $n$ is important because it determines the graphene Fermi momentum magnitude, $k_{F}=\sqrt{\pi n}$, and the corresponding Fermi energy, $\varepsilon _{F}=\hbar v k_{F}$, where $v=c/300$ is the Fermi velocity of the linearized band and c is the speed of light in free space. For intrinsic graphene ($n=0$), thus the Fermi energy $\varepsilon _{F}$ coincides with the Dirac point, i.e., we set $\varepsilon _{F} =0$. Similar scenario also describes the minigap $\Delta_l$ in other monoatomic conductors.

\subsection{Key parameters of the C/N-knot}
Basic parameters of the C/N-knot are deduced by measuring the thermal and electric currents. Besides, parameters are determined from the first-principle numeric calculations. In particular, for 2H-NbSe$_2$ ~\cite{Valla}, using the effective electron mass $m^*=0.6$~$m$ and lattice constants $a = 3.45$~\AA, $c = 12.54$~\AA, one evaluates the "bare" electron density of states (DOS) as $N(0) \simeq 2.8$ states/(eV$\cdot $ unitcell) and the Fermi velocity as $v \simeq 1.6 \times 10^6$~m/s. 
The electron minigap $\Delta _{m}$ arising in 2H-NbSe$_2$ due to presense of superlattice with a "long" period $L_{\rm SL} >> \max\{a,c\}$ is evalated as follows. Let us consider a wide range of densities, expressed as a multiple of the base value $n_{0}=10^{11}$~ cm$^{-2}$, under the conditions $k_{F}<k_{c}$ and $\varepsilon _{F}$ $<\varepsilon _{c}$. For the latter number, we evaluate the Fermi wavevector as $k_{F}=\sqrt{\pi n_{0}} =5.6\times 10^{7}$~m$^{-1}$, and $p_{F}=\hbar k_{F}=3.7 \times 10^{-8}$~eV s/m. For the superlattice period $L=50$~nm and the Fermi velocity $v=1.6 \times 10^{6}$~m/s, $K=2\pi /3a$, we find the minigaps $\Delta _{m} \simeq \left\vert f_{m}\right\vert \cdot 5.9 \times 10^{-2}$~eV spaced with $\Delta E_{s}=6.6\times 10^{-2}\text{ eV}$. Thus both, the minigap magnitude $\Delta _{m}$ and spacing $\Delta E_{s}$ are adjusted by changing the superlattice period and of amplitude of the "slow" potential $V(x)$. 
Similar subbands arising in the electron excitation spectrum of a narrow conducting stripe, or when applying a d.c. magnetic field considered in the following sections. 

\subsection{Nonequilibrium effects}
{\it Phonon drag} is essential in semiconducting stripes and nanotubes whereas it is negligible in metallic stripes and nanotubes. Phonon drag occurs~\cite{Bailyn,Bastard} when the non-equilibrium phonons propagating between the hot (C$^{\prime \prime }$) and cold (C$^{\prime }$) spots of the same C,  are generating the electrons and holes in the processes of phonon-electron collisions. In Eq.~(\ref{current}), the electron distribution function $f_{C} = f_{C}^{(0)} (T_{\mathrm{C}}^{\prime \prime}) + \delta f^{\mathrm{th}} + \delta f^{\mathrm{pd}}$ significantly deviates from its equilibrium value $f^{(0)} (T_{\mathrm{C}}) = 1/(\exp{(\varepsilon /T_{\mathrm{C}}}) +1) $. The non-equilibrium deviation $\delta f^{\mathrm{th}}$ is caused by two mechanisms. (a)~Due to the thermal injection from H to C$^{\prime \prime }$ by $\delta f^{\mathrm{th}}$ and (b)~due to the phonon drag generation by $\delta f^{\mathrm{pd}}$. The corresponding non-equilibrium deviations $\delta f^{\mathrm{th}}$ and $\delta f^{\mathrm{pd}}$ caused by (a) and (b) are computed using
Eqs.~(\ref{kin-el}) , (\ref{kin-eq}). The distribution function of electrons $f_{\mathrm{N}}^{\mathrm{left/right}}$ belonging to the left and right wings of the same N-stripe are approximately equal to each other (i.e., $f_{N}^{%
\mathrm{left}} \simeq f_{N}^{\mathrm{right}}$). Therefore, we merely regard them as equilibrium Fermi functions  $f^{\left( 0\right) }\left( \varepsilon/T_{\mathrm{C}} \right) =1/\left( \exp \left(\varepsilon /T_{\mathrm{C}}\right) +1\right) $ with a steady state  temperature $T_{\mathrm{C}}$. Equations~(\ref{kin-el})  with $\mathcal{L}_{th}$ were addressed in Ref.~\cite{Shafr-TEG0} whereas the phonon drag was considered in Refs.~\cite{Bailyn,Bastard,Shafr-TEG0,Shafr-TEG}. The phonon drag leads to the non-equilibrium deviation $\delta f^{\mathrm{pd}}$ which in the linear approximation is computed from Eq.~(\ref{kin-el}) as 
\begin{equation}
\delta f^{\mathrm{pd}}\simeq \pi \tau _{ep} g^{2}\sum_{m=\pm }\left\vert
m\left(\kappa_{m}\right) \right\vert ^{2}\delta N(s \kappa_m) [f_{m} + f^{\left(
0\right)}-2f^{\left( 0\right) }f_{m}]  \label{dfph}
\end{equation}
where $\tau _{\mathrm{ep }}$ is the electron-phonon collision time, $m=\pm 1$, $f_m \left( \varepsilon \right)=1/\left( \exp \left( \varepsilon_A^m\right) +1\right) $, $\kappa_m=2\left(k+m \left( s/v\right) \sqrt{k^{2}+q_{\nu }^{2}}\right) /\left( \left(
s/v\right) ^{2}-1\right) $, $q_{\nu }$ is the quantized transversal component of electron momentum in C, $\varepsilon_A^m = \sqrt{\varepsilon ^{2}+v^{2}\kappa_m^{2}+2 \kappa_m v \varepsilon_B}/T_{\mathrm{C^{\prime \prime }}} $, $\varepsilon_B = \sqrt{%
\varepsilon ^{2}-v^{2}q_{\nu }^{2}} $, $s$ is the speed of sound, and $v$ is the Fermi velocity in C. An independent linear deviation $\delta f^{\mathrm{th}}$ of the distribution function from its steady state $f_{\mathrm{C}}$ owing to the thermal injection  from H in C is evaluated as
\begin{equation}
\delta f^{\mathrm{th}} = \frac{\tau _{\mathrm{ep }} \Gamma _{\mathrm{HC}}}{1 + \tau _{\mathrm{ep }} \Gamma _{\mathrm{HC}}}(f_{\mathrm{C^{\prime \prime}}%
}-f^{(0)})  \label{dfth}
\end{equation}%
where $\Gamma _{\mathrm{HC}}$ is the rate of electron tunneling between H and C, $f_{\mathrm{C^{\prime \prime}}} = 1/(\exp{(\varepsilon /T_{\mathrm{C^{\prime \prime }}})}+1)$. In the above Eq.~(\ref{dfph}), $\delta N = N_{\mathrm{C}}-N^{(0)}$ is the non-equilibrium deviation of the phonon distribution function $N_{\rm C}$ from the Bose-Einstein form $N^{(0)} = 1/(\exp{(\omega /T_{\mathrm{C^{\prime }}})}-1)$ inside the C/N-knot is determined from Eq.~(\ref{kin-eq}). For the sake of simplicity we use a model form  
\begin{equation}
\delta N\left( \omega \right) = N_{0}e^{-\frac{\left( \omega -\Omega_{0}\right) ^{2}}{\sigma _{\mathrm{ph}}} } 
\label{del-N}
\end{equation}%
where the parameters $N_{0}$, $\Omega _{0}$, and $\sigma _{ph}$ are the amplitude, frequency, and width of the phonon beam propagating along C from C$^{\prime \prime }$ to C$^{\prime }$ respectively. The parameters are either determined from corresponding boundary conditions, or computed microscopically, and/or extracted from the experiments. The non-equilibrium thermal injection from H to C induces a beam of phonons (\ref{del-N}) along C. Intensity $N_0$ of the phonon beam (\ref{del-N}) inside C is determined with power of the heat source H, and with the H/C-interface transparency $T_{\mathrm{ph}}^{\mathrm{HC}}$. The beam width $\sigma_{\mathrm{ph}}$ depends on geometry of the source of heat H.
\begin{figure}
\includegraphics[width=3.5 in]{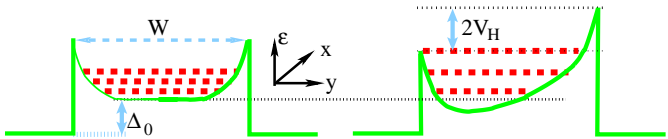}
\caption{Color online. Energy profile of the confinement potential  $U(y)$  formed at the C/N-knot in the transversal {\it \^y}-direction. (a)~$U(y)$ at ${\bf B} = 0$. (b)~The Hall voltage $2 V_{\rm H}$ tilts the $U(y)$-potential when a finite d.c. magnetic field ${\bf B} \neq 0$ is applied perpendicular the C/N-knot. Notice, that in the latter case,  the level spacing increases owing to the interplay between the Landau and spatial quantization.} 
\label{fig2s}
\end{figure}

\subsection{Thermal Hall effect}
The {\it thermal Hall voltage} $V_{\rm H}$ arises as follows. Let us consider the electron-hole symmetry assuming that the source of heat H injects the non-equilibrium electrons and holes into C in equal numbers. The finite difference $\delta T^{\ast }$ induces a drift of the excitations from C$^{\prime \prime}$ to C$^{\prime }$. The d.c. magnetic field deflects the propagating electrons and holes toward the opposite C-stripe edges. The C-stripe becomes electrically polarized in the $\hat y$-direction owing to the spatial separation of the electrons and holes. Thereby, the holes create an electric current along C near one edge whereas the current of electrons is localized on the opposite edge and flows in the opposite $\hat x$-direction. In the case of the electron-hole symmetry, the electric currents of electrons and holes, localized on opposite edges of C, mutually cancel each other, thus the total longitudinal electric current in C vanishes. Nevertheless, the transversal polarization of the electric charge creates a finite Hall voltage $V_{\rm H} \neq 0$ across the C-stripe. We emphasize that the above mechanism is quite different from the conventional Hall voltage requiring the electron-hole asymmetry when a certain type of charge carriers dominate. Besides, we assume that in vicinity of the C/N-knot, the Luttinger liquid~\cite{Kane-Fisher} is not formed due to coupling between the C and N electrodes. 
As electric charges propagate along the C-stripe, they experience the transversal Lorentz force $F_{M}=\pm ev_{d}B_{\perp }$ while moving in the magnetic field. Initially, the charges will move towards the edge of the stripe. As soon as the charges pile up across the C-stripe, they creates an electric potential between the edges of the stripe in the  transversal direction. Eventually, this potential will create a transversal electric force $F_{\rm E}$ on the charge carriers that will be equal to the magnetic force $F_{\rm M}$ on them. The system will then reach a quasi-stationary state in which there will be a constant voltage $V_{H}$ across the stripe. In the quasi-stationary state, provided that $F_{E} =F_{M}$ and $eE = ev_{d}B_z$, the constant currents of electrons and holes are getting localized on the opposite C-stripe edges. The transversal electric field $E$ created by the pile up of charges is constant in the stripe. The $E$-field is equal to the voltage $V_{\rm H}$ across the stripe divided by the width $W$ giving $E=V_{\rm H}/W$. The drift velocity $v_{d}$ of the charge carriers and the heat flow $Q$ are related to each other as follows. The amount of heat transferred due to the drift of non-equilibrium electrons between the C$^{\prime \prime }$ and C$^{\prime }$ sections per unit time is 
\begin{equation}
Q_{\rm ne}=\left( \frac{\bar{\varepsilon}_{el}}{\Delta t}\right) L_{C} W h=\bar{\varepsilon}_{el}\left( \frac{L_{C} }{\Delta t}\right) Wh =\bar{\varepsilon} _{el} v_{d}W h
\end{equation}%
where $L_{C} = v_{d} \Delta t$ is the spacing between the C$^{\prime \prime }$ and C$^{\prime }$ sections, $\Delta t $ is the drift time, $h$  is the stripe thickness.
The energy of non-equilibrium electrons per unit volume is obtained as
\begin{equation}
\bar{\varepsilon} _{el}=\int d\varepsilon M\left(\varepsilon \right) (\delta f^{\mathrm{pd}} + \delta f^{\mathrm{th}})\cdot \varepsilon .
\label{energy}
\end{equation} 
The non-equilibrum deviations $\delta f^{\mathrm{pd}}$ and $\delta f^{\mathrm{th}}$ of the electron distribution functions are given by Eqs.~(\ref{dfph}), (\ref{dfth}) correspondly. On the other hand, from the Fourier law, we find $Q_{\rm F}=\Lambda _{el}\left( T^{\ast }-T_{C}\right)$. Using $Q_{\rm ne} = Q_{\rm F}$, we obtain the drift velocity as $v_{d}=\Lambda _{el}\left( T^{\ast }-T_{C}\right) /(\bar{\varepsilon}_{el} W h)$.
With allowance of the above equations, we obtain the Hall voltage as
\begin{eqnarray}
V_{H} &=&\frac{\Lambda _{el}\left( T^{\ast }-T_{C}\right) }{\bar{\varepsilon} _{el}h}B_z.
\label{VH}
\end{eqnarray}%
which enters Eq.~(\ref{current}). We emphasize that the finite Hall voltage $V_{\rm H} \neq 0$ causes a build up of the electric potential only in C, thereby creating the bias voltage between the C and N stripes, whereas the electric field along N vanishes (see Fig.~\ref{fig2s}).

\end{document}